# Sampling molecular conformations and dynamics in a multi-user virtual reality framework

Michael O'Connor,[1,2,3,‡] Helen M. Deeks,[1,2,‡] Edward Dawn,[1] Oussama Metatla,[2] Anne Roudaut,[2] Matthew Sutton,[2] Becca Rose Glowacki,[3,4] Rebecca Sage,[3,5] Philip Tew,[3,5] Mark Wonnacott,[5] Phil Bates,[6] Adrian J. Mulholland,[1] and David R. Glowacki[1,2,3*]

[1]Centre for Computational Chemistry, School of Chemistry, University of Bristol, Cantock's Close, Bristol BS8 1TS, UK; [2]Dept. of Computer Science, University of Bristol, Merchant Venturer's Building, Bristol BS8 1UB, UK; [3]Pervasive Media Studio, Watershed, 1 Canons Rd, Bristol BS1 5TX, UK; [4]School of Art & Design, University of the West of England, Coldharbour Lane, Bristol BS16 1QY, UK; [5]Interactive Scientific Ltd., Engine Shed, Station Approach, Bristol BS1 6QH, UK; [6]Oracle Cloud Development Centre, Tower Wharf, Cheese Lane, Bristol, BS2 2JJ, UK

‡ These authors contributed equally to this work
*glowacki@bristol.ac.uk

*We describe a framework for interactive molecular dynamics in a multiuser virtual reality environment, combining rigorous cloud-mounted atomistic physics simulations with commodity virtual reality hardware, which we have made accessible to readers (see isci.itch.io/nsb-imd). It allows users to visualize and sample, with atomic-level precision, the structures and dynamics of complex molecular structures 'on the fly', and to interact with other users in the same virtual environment. A series of controlled studies, wherein participants were tasked with a range of molecular manipulation goals (threading methane through a nanotube, changing helical screw-sense, and tying a protein knot), quantitatively demonstrate that users within the interactive VR environment can complete sophisticated molecular modelling tasks more quickly than they can using conventional interfaces, especially for molecular pathways and structural transitions whose conformational choreographies are intrinsically 3d. This framework should accelerate progress in nanoscale molecular engineering areas such as drug development, synthetic biology, and catalyst design. More broadly, our findings highlight VR's potential in scientific domains where 3d dynamics matter, spanning research and education.*

It is a fundamental human instinct to build and manipulate models to understand the world around us. Chemists have constructed three-dimensional molecular models and used them as conceptual and educational tools, dating back to at least von Hofmann in the 1860s. The use of tangible molecular models has a long history within both chemistry and biochemistry. Influential examples include Dorothy Hodgkin's crystallographic model of penicillin's structure, (*1*) demonstrating the presence of a β-lactam ring, and Pauling's use of (originally paper!) models to identify the structure of alpha-helices. (*2*) Perhaps the best-known model is captured in the iconic photo of Watson and Crick discussing their physical model of DNA. (*3*) Large room-sized models, made from e.g., wire, plastic, brass, balsawood, and plasticene were used to refine and represent protein crystal structures by pioneers such as Kendrew and Perutz (*4, 5*). For example, Michael Levitt (in his 2013 Noble lecture) recounted building a so-called "Kendrew Model" of hen egg white lysozyme: "…painful slow work but at the end you really know the molecule." (*6*) Nowadays, solid models can be fabricated by 3-D printing. (*7*)

Physical models like these provide insight, but cannot represent the mechanics that determine how molecules move and flex. Scientists have since made heroic efforts to include the fourth dimension (time) into their models. Simulation of molecular motion has its origin in the work of pioneers like Bernal, whose attempts to understand liquid structure involved mechanical simulation of assemblies of macroscopic spheres. (*8*) In the 1970s, pioneers like Karplus, Warshel, and Levitt were able to simulate the motion of complex molecules such as proteins. 1979 saw the first movie of a molecular dynamics (MD) trajectory (bovine pancreatic trypsin inhibitor), (*9*) one of several developments heralding the replacement of tangible mechanical models by screen-based graphics. These rose to prominence in the 1980s and are now ubiquitous for the purpose of understanding and teaching molecular structure. (*10*)

Advances in computer power, and decreasing cost, have transformed molecular visualization and simulation. For example, graphical processing units (GPUs), designed to facilitate the fast rendering required by modern video games, have not only benefitted scientific visualization; they have also been adapted to accelerate a range of molecular physics algorithms, (*11*) now allowing routine simulation of systems with hundreds of thousands of atoms at timescales ranging from hundreds of nanoseconds to microseconds, with applications to areas including protein folding, (*12*) enzymology, (*13*) and drug discovery. (*14*) Beyond GPUs, application-specific integrated circuits (ASICS), (*15*) cloud computing platforms, (*16*) distributed computing networks, (*17*) and emerging architectures enable simulation of large systems at millisecond timescales. (*15, 16*) Nevertheless, chemical transformation is characterized by considerable complexity, in part because it involves rare events in hyper-dimensional systems. There is good evidence that many such transformations belong to a class of "NP-complete" problems, for which no obviously optimal solution exists. (*18*) For the foreseeable future, many important molecular-level transformations occur on timescales that will remain beyond even the most sophisticated simulation architectures. Chemical "intuition" (an oft-invoked concept that broadly refers to the chemist's ability to efficiently navigate hyperdimensional molecular space) is therefore likely to play an important role in (bio)chemistry and synthetic biology for a long time to come. Tools accelerating the rate at which chemical intuition can be brought to bear on structurally and physically detailed models will facilitate creative teams making progress on challenging problems.

For the last 30 – 40 years, the basic workflow for molecular simulation has (like many domains of scientific computing) remained largely unchanged – i.e., iterative cycles of job submission to HPC resources, followed by visualization on a 2d display. Pioneers with interests spanning molecular simulation and human-computer-interaction (HCI), including Fred Brooks (*19*) and Kent Wilson (*20*), were amongst the first to imagine improvements to this workflow using interactive computational technologies: they speculated that interactive molecular simulation (iMS) frameworks would lead to models as intuitive to manipulate as the old tangible models, but which followed rigorous physical laws, and which could be used to tackle hard rare event sampling problems. Brooks designed an immersive six-degree-of-freedom force-feedback haptic system which users could manipulate to carry out simple molecular docking tasks. (*19, 21*) Inspired by this work, Klaus Schulten and co-workers subsequently miniaturized Brooks' setup:

by manipulating a desktop-mounted haptic pointer, users could steer the real-time dynamics of molecules rendered on a stereographic screen, (*22*) a setup which has since been extended by others [e.g., (*23-25*)].

These interactive setups face a well-known limitation in their ability to achieve 3d 'co-location' – i.e., aligning the interaction sites in 3d physical space and 3d virtual space. (*26*) Touchscreens solve the problem of 2d co-location because the interaction site in physical space is identical to the virtual interaction site. For interactive molecular simulations, 3d co-location is an important design problem, given that molecules are 3d objects that move in 3d. Despite these difficulties, the utility of the iMS idea has been demonstrated by tools like Foldit. (*27*) Using a keyboard-mouse interface, Foldit enables users to apply their intuition to explore protein conformational space and predict protein structures. Cooper *et al.* highlighted cases where Foldit users made better predictions (i.e., located deeper energetic minima) than automated algorithms, owing to users' willingness to explore high-energy ("high risk") pathways which automated strategies avoided. Khatib *et al.* analyzed user search strategies to construct new algorithms. (*28*)

Driven by demands of the consumer gaming market, recent advances in VR hardware provide commodity-priced solutions to the problem of 3d co-location. Combining infrared optical tracking, inertial movement units (IMUs), and ASICS, high-end commodity VR technology like the HTC Vive tracks a user's real-time 3d position with errors less than a centimeter. The bottom panel of Fig 1 shows the framework we have developed to interface the HTC Vive with rigorous real-time MD simulation algorithms. (*29*) Two optically tracked researchers are shown (each wearing a VR head-mounted display (HMD) and holding two small wireless controllers which function as atomic 'tweezers') manipulating the real-time MD of a $C_{60}$ molecule. As shown in **Supplementary Video 1,** the researchers can easily 'grab' individual $C_{60}$ atoms, and manipulate their real-time dynamics to pass the $C_{60}$ back and forth between each other. This is possible and immediately intuitive because the real-time $C_{60}$ simulation and its associated ball-and-stick visual representation is perfectly co-located – i.e., the interaction site in 3d physical space is exactly the interaction site in 3d simulation space. The cloud architecture provides further benefits: because each VR client has access to global position data of all other users, any user can see through his/her headset a co-located visual representation of all other users at the same time. To date, our available resources have allowed us to simultaneously co-locate six users in the same room within the same simulation. The framework also allows users who are not physically co-located to work together in the same virtual 'room', facilitating collaboration.

While a real-time $C_{60}$ simulation is relatively cheap, the cloud architecture enables access to more powerful computational servers as needed. For example, the top panel of Fig 1 and **Supplementary Video 2** show a researcher taking hold of a fully solvated benzyl-penicillin ligand and interactively guiding its dynamics to dock it within the active site of the TEM-1 β-lactamase enzyme and generate the correct binding mode, (*30*) a dynamical process important to our understanding of anti-microbial resistance. (*31*) The β-lactamase example (which benefits from a plugin that communicates with OpenMM (*32*) via PLUMED (*33*)) illustrates that our framework is sufficiently intuitive and easy-to-control so as to enable a researcher to quickly formulate and evaluate dynamical hypotheses in large molecular systems. Generating a benzyl-penicillin docking pathway involves a non-linear sequence of complex molecular manipulations that would be difficult to formulate algorithmically. The framework thus has potential as an effective tool for understanding molecular docking and kinetics, standard tasks in structure-based drug design – and may prove particularly effective as a complement to enhanced sampling techniques for challenges such as identifying allosteric and cryptic binding sites that are not present in crystal structures. (*34*)

An essential question is whether co-located interaction of the sort enabled by VR affords any real advantage in accelerating complex 3d manipulation tasks compared to more conventional interaction technologies. To tackle this question, we exploited the fact that the Fig 1 framework runs on a wide range of different client interaction hardware, including mice (connected to PCs/laptops) and tablets. Both are extremely familiar interfaces, the former offering non-co-located interaction, and the latter offering co-located interaction in 2d. We designed three specific tasks (Fig 2 and **Supplementary Video 1**) for users to carry out, and compared the rates of task accomplishment across different interaction hardware. Each task involves an increasingly complicated choreography: (1) guiding $CH_4$ through a carbon nanotube; (2) changing the screw-sense of an organic helicene molecule; and (3) tying a knot in a small 17-ALA polypeptide. Each task represents a class of dynamics which is important across natural and engineered nano-systems. For example, nano-pores are ubiquitous across bio/materials chemistry; (*35*) induced changes in molecular helicity offer a strategy for synthetic biologists to transmit chemical messages; (*36*) and protein knots are associated with neurodegenerative diseases. (*37*) To control for variations in the graphics and computational capabilities of different interaction hardware, each interaction platform utilized the same renderer, and the same back-end molecular dynamics simulation engine. The SI includes details on the simulation setups, along with instructions enabling readers to use the executables at *isci.itch.io/nsb-imd* to initialize a cloud-based interactive simulation instance so that they can attempt the Fig 2 tasks on any of the three interaction platforms.

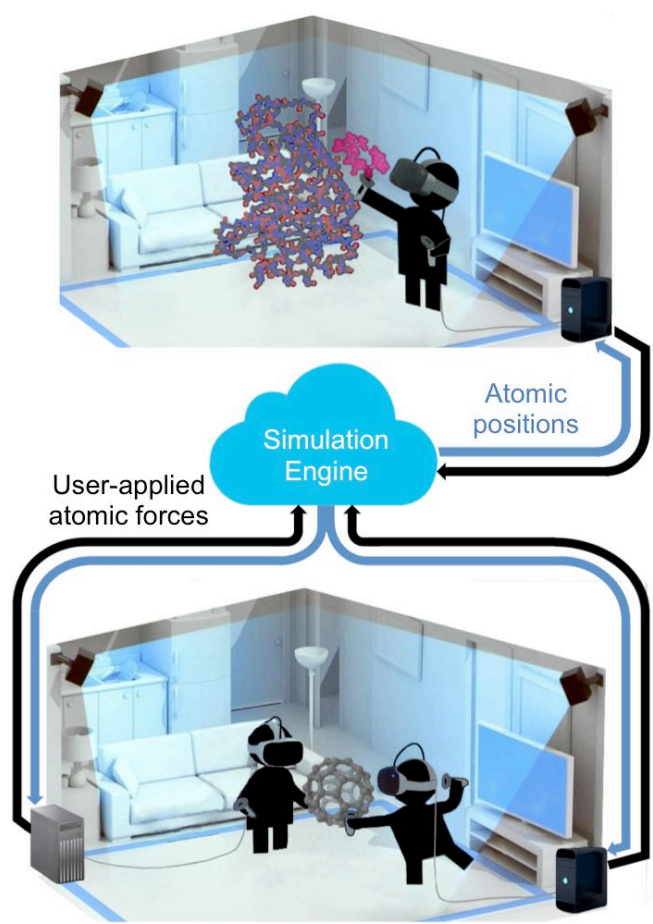

Figure 1: technical schematic of the HTC Vive VR setup we designed to carry out the studies outlined herein. Bottom panel shows two users within the multi-person VR framework passing a simulated $C_{60}$ molecule back and forth. Each user's position is determined using a real-time optical tracking system composed of synchronized IR light sources. Each user's HMD is rendered locally on a computer fitted with a suitable GPU; molecular dynamics calculations and maintenance of global user position data take place on a separate server, which can be cloud-mounted. So long as the network connecting client and server enables sufficiently fast data transfer, system latency is imperceptible to the human senses. The top panel shows a single-person setup, where the user is chaperoning a real-time GPU-accelerated MD simulation to generate an association pathway that docks a benzyl-penicillin ligand (magenta) into a binding pose on the TEM-1 β-lactamase enzyme.

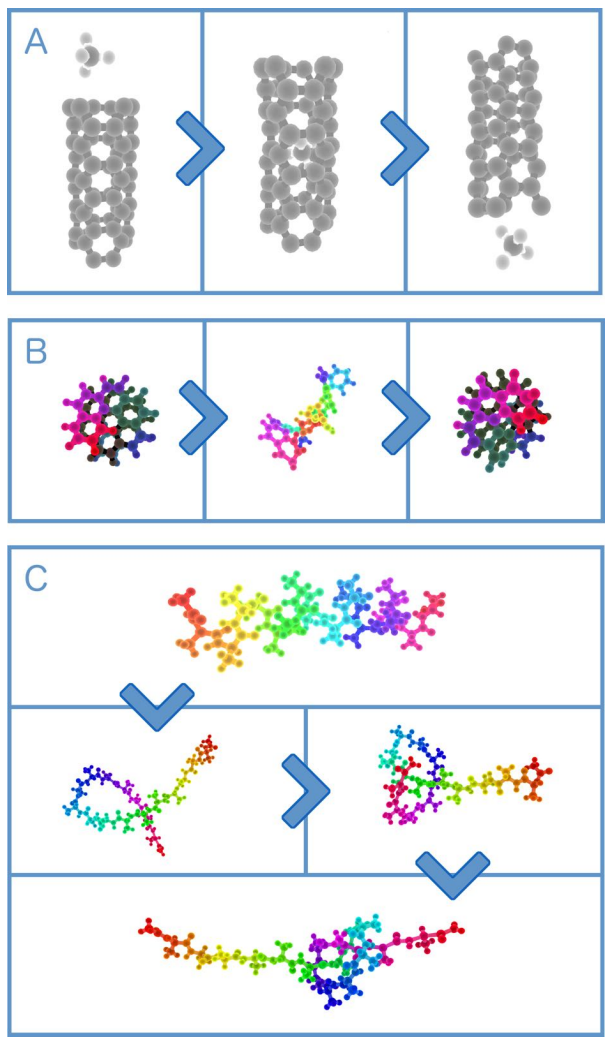

Figure 2: interactive molecular simulation tasks used as application tests: (1) threading $CH_4$ through a nanotube; (2) changing the screw-sense of a helicene molecule; and (3) tying a knot in a polypeptide. Colors selected in this figure are chosen for the sake of clarity.

Fig 3 shows the rates at which cohorts of 32 users (who reported little or no previous VR experience) accomplished each of the three tasks in Fig 2 on different hardware platforms. Prior to their participation, study candidates were warned of the potential for VR sickness, and instructed to inform the demonstrators if they wanted to exit at any point. We recorded zero such instances, suggesting that our design produces little user discomfort. Reported errors in the measured rates were obtained using Poisson statistics, a common strategy in chemical kinetics: rare event (or 'task accomplishment') rates are said to be statistically distinguishable if their corresponding Poisson error bars do not overlap. For tasks A and C, Fig 3 indicates that VR provides a clear acceleration benefit. The more inherently 3d the task, the greater the benefit, with the knot-tying results (Fig 3c) the most dramatic: our show that this task is very difficult to accomplish using a mouse or a touchscreen. To ascertain that the knot-tying results were in fact reproducible, we carried out another set of tests with a different cohort; reassuringly, the results were the same within Poisson error limits (see SI). For the nanotube task (Fig 3a), the accomplishment rates, mean time, and median time in VR are approximately a factor of two faster than on other platforms.

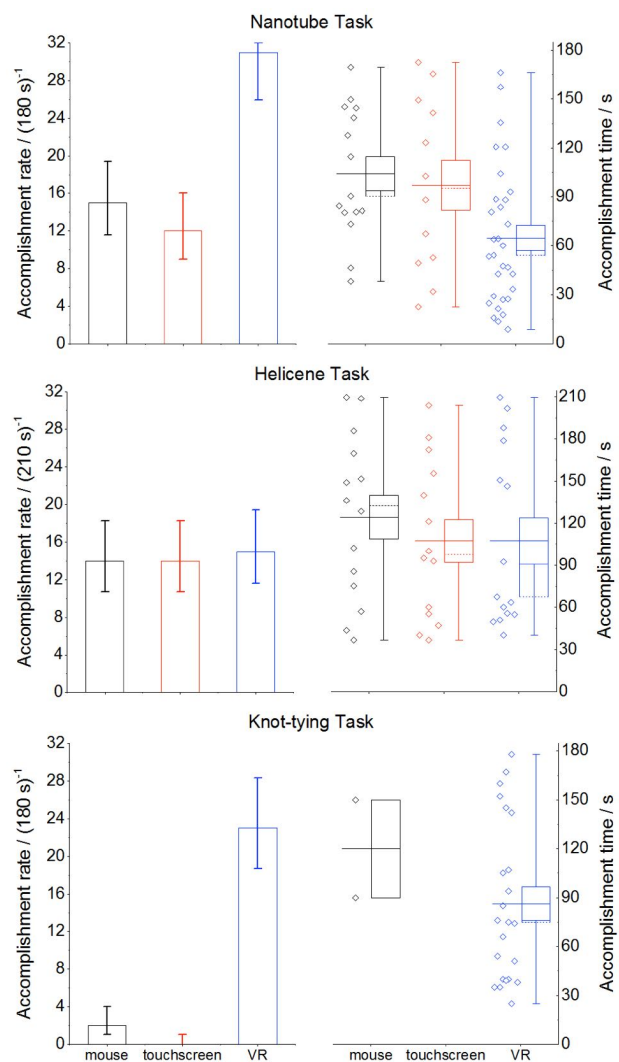

Figure 3: User study results. Left-hand panel shows user accomplishment rates for the tasks outlined in Fig 2 (n = 32 for all tasks), with Poisson error estimates. Right hand panel shows the corresponding distribution of task accomplishment times, along with box-and-whisker plots. Whiskers indicate the data range and box limits the standard error of the distribution. The mean is shown as a solid line, and the median as a dashed line.

At first glance, the helicene task (Fig 3b) is a case for which VR appears to provide little significant rate enhancement compared to other platforms. Observation suggests that this is because changes in helicene screw-sense are most efficiently accomplished using a simple 2d circular motion, as shown in **Supplementary Video 1**. Essentially, the 2d limitations of the mouse and touchscreen constrain the user to carrying out a motion which is well suited to inducing changes in molecular screw-sense. However, closer inspection of the helicene time distributions suggest that VR does afford some advantage: the median time required to change molecular screw-sense in VR is 30–40% less than the median time required on a touchscreen. Significantly, every task considered in Figs 2 and 3 shows a median VR time which (i) is always faster than the mean, and (ii) lies outside the standard error limits. This distribution of waiting times is clearly not Gaussian; rather it resembles the waiting time distribution that characterizes first order kinetics, suggesting a cluster of VR users who accomplish tasks quickly with a longer tail of slow users.

Fig 3 shows that our VR platform provides a clear advantage over conventional visualization/interaction platforms, and this advantage appears to increase as the complexity of the problem increases. To better understand the results, we carried out further evaluation in the form of participant questionnaires and interviews (detailed in the SI). Despite the fact that very few of the study participants had prior VR experience, participants indicated an overwhelming preference for VR in carrying out the Fig 3 tasks (compared to a mouse or touchscreen) for three reasons: (i) quick perception of depth; (ii) the ability to inspect the molecule by simply walking around it, and (iii) the use of both hands to accomplish tasks.

The results presented herein represent a first attempt to quantify the acceleration which can be achieved by carrying out 3d molecular sampling, modelling, and manipulation tasks in VR. Molecules offer particularly interesting candidates for investigating VR's potential because (i) they are flexible objects with many degrees of freedom and therefore complicated dynamics; and (ii) their physics is reasonably well-characterized. For biomolecular conformational problems like dynamical path sampling or drug docking – where computational search spaces are too large for brute force approaches – our results suggest that VR frameworks like those which we have outlined herein may offer a powerful tool enabling research experts (e.g., structural biologists, molecular modelers, medicinal chemists) to occupy the same virtual environment and efficiently express their 3d biomolecular intuition to test hypotheses for collaboratively tackling important microscopic questions linked to molecular mechanism and design. For example, using VR technologies to generate seeds for subsequent automated searches could represent a powerful new strategy for obtaining molecular insight, which is likely to save on valuable computational clock cycles. It may even be possible to imagine a scenario where machines 'learn' efficient strategies for navigating hyper-dimensional 3d search spaces by accumulating observational data on how human experts navigate hyper-dimensional biomolecular spaces in VR. In future studies, we aim to further investigate VR's potential for advancing molecular science in a range of research, communication, and educational contexts, and study its ability to enable specialists and non-specialists alike to better understand and more efficiently manipulate complex data.

**Acknowledgments.** MOC, PT, MW, and RS designed and implemented the cross-platform, real-time, cloud-mounted multi-person iMD framework. HMD and ED carried out user studies and performed data analysis. HMD, OM, and AR designed the user studies. MS constructed the video figures. BRG and DRG designed the molecular tasks. PB provided crucial support implementing the cloud-mounted simulation infrastructure. DRG designed the overall project concept, organized execution of the work strands, analyzed the data, and wrote the initial paper draft along with HD and MOC, with input from AJM, BRG, and AR. We also thank the following individuals: Dr. Thomas Mitchell, Prof. Joseph Hyde, Lisa May Thomas, Isabelle Cressy, Benjamin de Kosnik, Gemma Anderson, and Rob Arbon, all of whom contributed during the Barbican Open Lab residency where the aesthetics of the multi-person VR environment took shape; Richard Male (UFI Charitable Trust), Anna Wilson (Barbican), Sidd Khajuria (Barbican), Chris Sharp (Barbican), James Upton (Royal Society), Simon McIntosh-Smith (University of Bristol [UoB]]), and Dek Woolfson (UoB) for encouragement and support throughout; Joseph A. Glowacki (UoB) for assisting HMD with data collection; Gerardo Viedma Nunez and Jenny Tsai-Smith (Oracle) for guidance in enabling the cloud-mounted simulation engine; Jono Sandilands and Liam Cullingford (Interactive Scientific) for respective cloud-mounted interaction design contributions and project management; and the 2017 CCP-BioSim annual conference and its organizers (in particular Dr. Marc van der Kamp and Prof. John Essex) for enabling us to run user studies as part of the conference program (ccpbiosim.ac.uk). MOC's studentship (supervised by DRG) is supported through an industrial CASE award, funded by EPSRC and Interactive Scientific. HMD's PhD studentship (supervised by DRG, AJM, HD and OM) is funded by EPSRC. DRG acknowledges additional funding from: Oracle Corporation (University Partnership Cloud award); the Royal Society (UF120381); EPSRC (impact acceleration award, institutional sponsorship award, and EP/P021123/1), the Leverhulme Trust (Philip Leverhulme Prize); and the London Barbican (Open Lab Funding). BS and PT acknowledge support from the UFI Charitable Trust. AR acknowledges support from the Leverhulme Trust and EPSRC (EP/P004342/1). AJM thanks EPSRC for funding (EP/M022609/1).

# Supplementary Information

Michael O'Connor,[1,2,3,‡] Helen M. Deeks,[1,2,‡] Edward Dawn,[1] Oussama Metatla,[2] Anne Roudaut,[2] Matthew Sutton,[2] Becca Rose Glowacki,[3,4] Rebecca Sage,[3,5] Philip Tew,[3,5] Mark Wonnacott,[5] Phil Bates,[6] Adrian J. Mulholland,[1] and David R. Glowacki[1,2,3*]

[1]*Centre for Computational Chemistry, School of Chemistry, University of Bristol, Cantock's Close, Bristol BS8 1TS, UK;* [2]*Dept. of Computer Science, University of Bristol, Merchant Venturer's Building, Bristol BS8 1UB, UK;* [3]*Pervasive Media Studio, Watershed, 1 Canons Rd, Bristol BS1 5TX, UK;* [4]*School of Art & Design, University of the West of England, Coldharbour Lane, Bristol BS16 1QY, UK;* [5]*Interactive Scientific Ltd., Engine Shed, Station Approach, Bristol BS1 6QH, UK;* [6]*Oracle Cloud Development Centre, Tower Wharf, Cheese Lane, Bristol, BS2 2JJ, UK*

*‡ These authors contributed equally to this work*
*\*glowacki@bristol.ac.uk*

This Supplementary Information includes the following sections:
1. Specific details on the molecular dynamics simulation setups which we used to carry out the user studies
2. Instructions for launching a cloud hosted interactive MD session
3. Details on user study design, along with additional data obtained during those studies
4. Web URLs for the Supplementary Videos
5. References

# 1. Molecular Dynamics Simulations

The broad strategies which we have utilized to carry out interactive MD are outlined in ref (*1*), and briefly summarized here for the sake of completeness. In classical mechanics, the time-dependent dynamics of molecular systems are solved by numerically integrating Newton's equations of motion. The vector of forces acting on a set of atoms **F**(*t*) can be written in terms of the system's potential energy *V*, i.e.:

$$\mathbf{F}(t) = -\frac{dV}{d\mathbf{q}} \qquad (S1)$$

where **q** is a vector containing the position of each atom in the ensemble. Our system effectively allows users to interactively chaperone a real time MD simulation by splitting *V* into two different components

$$V = V_{int} + V_{ext} \qquad (S2)$$

where $V_{int}$ corresponds to the system's internal potential energy, and $V_{ext}$ corresponds to the potential energy which a user exerts on a specific atom (or group of atoms) when he/she grabs it using the handheld wireless controller shown in Fig 1 of the main text. Substituting Eq (S2) into Eq (S3) then gives

$$\mathbf{F}(t) = -\frac{dV_{int}}{d\mathbf{q}} - \frac{dV_{ext}}{d\mathbf{q}} \qquad (S3)$$

The external forces are implemented by projecting a spherical Gaussian field into the system at the point specified by the user, and applying the field to the nearest atom *j* through the following formula

$$\frac{dV_{ext}}{d\mathbf{q}} = \frac{m_j c}{\sigma^2}(\mathbf{q}_j - \mathbf{g_i}) e^{\frac{-\|\mathbf{q}_j - \mathbf{g_i}\|^2}{2\sigma^2}} \qquad (S4)$$

where $m_j$ is the atomic mass of the atom, $c$ is a scale factor that tunes the strength of the interaction, $\mathbf{q}_j$ is the position of atom $j$, $\mathbf{g}_i$ is the position of the interaction site, and $\sigma$ controls the width of the interactive fields. For all tasks on all of the platforms, $c$ was set to 2000 (kJ*mol$^{-1}$)/(a.m.u.), a value that achieves responsive interaction while preserving dynamical stability, and $\sigma$ was set to the default value of 1nm. While an interaction is active, it is always applied to the same atom, which means a user can dynamically adjust the course and strength of the interaction simply by repositioning their field with respect to the atoms with which they are interacting, until he/she decides to 'let go'. As discussed in the main text, the position of the interaction field in VR is co-located to precisely where the user's controller is; for the mouse and touchscreen interfaces, the field is attached to the nearest atom (measured in the 2d plane of the screen). Movements of the field are only possible in 2d; so 3d motions must be built up from successive 2d motions followed by repositioning of the camera.

The simulation of internal forces depends on the specific system under investigation, and here we benefit from the fact that our framework has been designed to flexibly communicate with a wide range of force engines via a defined application programming interface (API). For the hydrocarbon systems present in the nanotube and helicene task, our own 'in-house' implementation of the MM3 forcefield was used. (*2*) The 17−Alanine protein knot task used the GPU-accelerated Amber99SB-ILDN forcefield provided within the OpenMM molecular dynamics package (*3*). To integrate the forces, all tasks used a Velocity Verlet integrator, with a Berendsen thermostat (*4*) set to a target temperature of 200K. For the nanotube and helicene tasks, a time step of 1fs was used, while the protein knot task used a time step of 2fs along with the RATTLE holonomic constraint, owing to the fact that the conformational change occurs over a longer timescale.

To carry out the user studies (detailed further in what follows), the simulations were hosted on separate machines within our own local area network. We used one machine for each task, in order to avoid circumstances where latency could arise from excessive computational load on single machine. The three machines that we used as simulation servers during user studies included the following: (1) a mid-range gaming desktop with a 3.5GHz quad-core Intel i5-6600K processor and a Nvidia GTX 970 graphics card; (2) a high-end Alienware13 R3 gaming laptop with a 2.6GHz quad-core Intel Core i7-6700HQ processor and an Nvidia 1060 dedicated VR graphics card; and (3) a high-end Alienware15 R3 gaming laptop with a 2.6GHz quad-core Intel Core i7-6700HQ and an Nvidia 1070 GTX dedicated VR graphics card. The mouse tests described below in section 3.1 (run at our lab in the Bristol chemistry department) were carried out using machine (1) along with an external USB mouse and a 21" external monitor. The mouse tests described below in section 3.3 (run at the CCP-Biosim 2017 conference) were carried out using the screen on the Alienware15 along with an external USB mouse. During all user studies, we throttled the performance on all platforms in order to guarantee that (despite differences in each machine's specifications) they were capable of running at 30 FPS. This ensured that the latency across all test platforms gave an equally fluid user experience. For the touchscreen version of the tasks, we used a Samsung Galaxy S3 tablet, connected to the simulation over an 802.11ac Dual Band Gigabit wifi connection.

## 2. Launching a cloud hosted interactive MD session

Binaries for each platform (Windows desktop, Windows with SteamVR, Mac OS X, and Android) are available at https://isci.itch.io/nsb-imd

To run a task on a particular piece of interaction hardware, simply download the appropriate version of the application for your device and launch. Note that Android users may need to edit their security settings to allow installation of apps from external sources. Follow the in-app buttons to select from one of the available servers which will host the simulations, then choose a molecular task to try. The following devices and operating systems are supported:

- Desktop/laptops running Windows XP SP2 or higher; or Mac OS X 10.9 or higher;
- Android touchscreen devices running Android OS 4.1 or later; with either an ARMv7 CPU with NEON support or Atom CPU, and OpenGL ES 2.0 or later.
- For the VR version of the app, a desktop running Windows 10 with a VR-capable GPU, as well as an HTC Vive headset and controllers.

Note that iOS is not currently supported. We have tested the application on a variety of machines, including Windows 10 desktops; Macbook Pro laptops running OS X High Sierra; Samsung Galaxy phones (S5, S6, S7) and tablets (S3); and Google Pixel, Pixel XL and Nexus 5 phones. Any issues or bugs found when using supported devices should be reported by leaving a comment on the itch.io website.

The latency of cloud-mounted interactive simulations depends on network speeds and wifi performance. Note that the experience using the application will vary according to one's distance from the server. For best results, we recommend choosing your simulation server to be hosted on the Oracle data center (Frankfurt, Germany; Phoenix, Arizona; Ashburn, Virginia) nearest to your physical location. Round trip measurements of the average latency (± standard deviation) for transmitting data from our lab in Bristol to each of the three Oracle data centers were as follows: 48.7 ± 9.5 ms for Bristol to Frankfurt; 118.7 ± 11.5 ms for Bristol to Ashburn; and 182.1 ± 18.3 ms for Bristol to Phoenix. For each round trip measurement, we investigated performance when interaction and visualization clients were connected via wifi (the Bristol eduroam network), and also via an ethernet cable. The results are shown in Fig S1. The latencies measured on wifi were statistically indistinguishable from those measured on an ethernet connection. Outside the confines of the lab, we noticed that we obtained acceptable latencies even on 4G mobile networks.

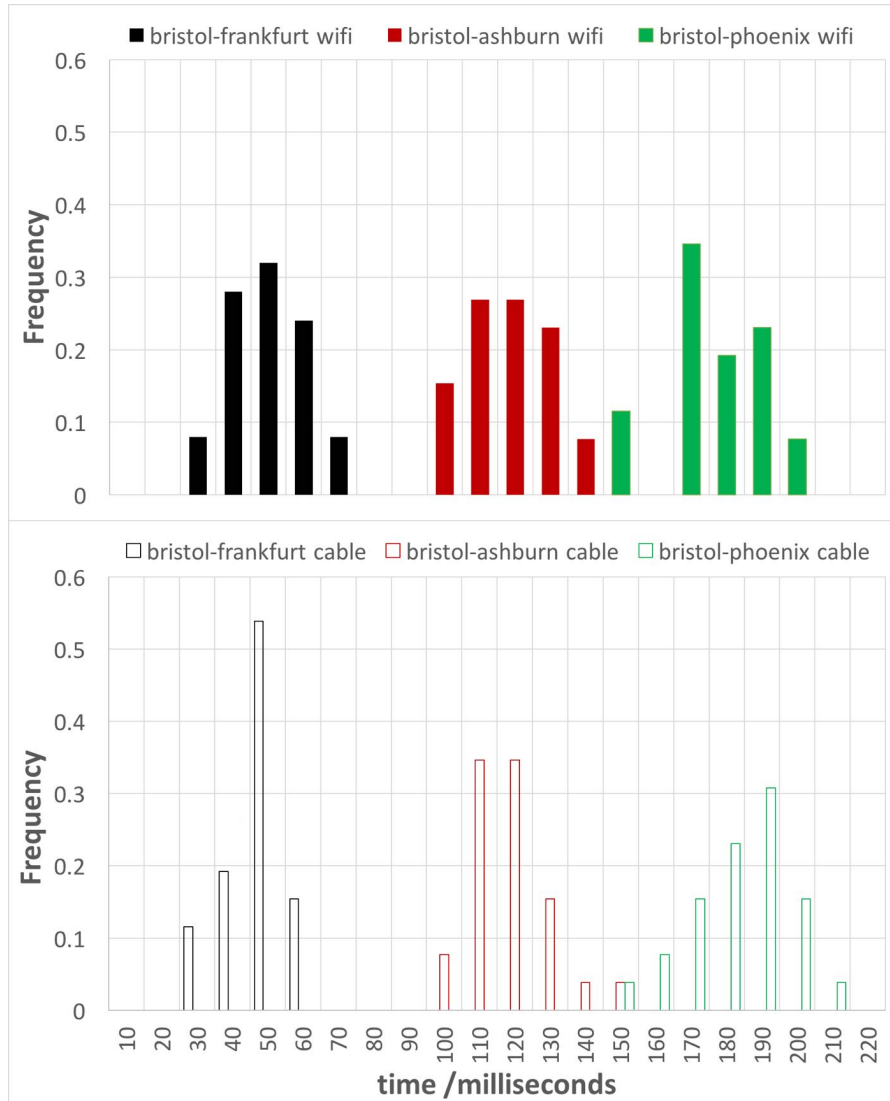

Figure S1: distribution of round-trip latencies, measured from Bristol to each of the cloud data centers. The top panel shows measurements where the interaction/visualization client is connected via wifi, and the bottom panel shows measurements where the connection was via ethernet cable.

As described above, the simulation servers utilized during our user studies were hosted on machines operating on a local network. The reason for this is that, at the time this work was carried out, Oracle's Frankfurt-based European cloud resources which give the best latencies from Bristol (as shown in Fig S1) were not yet available. The cloud-mounted simulation engine offers the most straightforward way for interested readers to test our framework, and also eliminates a host of issues which could arise from hardware and OS incompatibilities. Those readers who are unable to achieve reasonable latency on the available cloud servers but who wish to test out the framework described herein should contact the corresponding author to obtain a software testing license.

## 3. User Study Design

The results outlined in the main text were obtained from three separate user studies, each of which is described in further detail below.

*3.1 First User Study*

A total of 32 participants were recruited for this study. In order to mitigate any learning or fatigue effect, the platform on which any given participant started was randomly selected. 12 participants began with the mouse platform, nine participants started with the touchscreen platform, and 11 started with the VR platform. Participants were rotated using a Latin Square in the order of VR, mouse, touchscreen.

Before starting on a platform, participants were given the buckminsterfullerene task (see section 3.4 for details) in order to familiarize themselves with the 'feel' of the molecular interaction on a given platform. Once participants indicated to the study facilitators that they had a sufficient level of familiarity, they were moved onto the nanotube/methane task (see section 3.4 for details). Study facilitators moved a participant onto the helicene task if they either managed to accomplish the nanotube task, or time expired (see section 3.4 for details). Once the user had attempted both tasks, the process was repeated on the next platform until the participant had tried the task on all three platforms. Once participants had attempted both tasks on all three platforms, they were given a short questionnaire to fill out, details of which are discussed in further detail below.

In total, 32 participants were recruited through email to staff and students at the University of Bristol, and offered a £10 Amazon gift voucher for their time. 17 (53.125%) of the participants were ages 18 to 24, 10 (31.25%) were ages 25 to 34, four (12.5%) were ages 35 to 44, and one (3.125%) was ages 45 to 54. 22 (68.75%) of the participants were male. Participants reported a range of education levels. 11 (34.375%) were undergraduate students, 16 (50%) participants were postgraduate students, three (9.375%) were post-doctoral researchers, and two (6.25%) were researchers.

Participants were given a Likert scale question to complete in order to indicate their familiarity with using VR and tablets, where 1 represents having no experience and 5 represents being very experienced. A breakdown of responses can be found in Table S1. Altogether, self-reported VR experience was found to be low, where tablet use was more prevalent amongst the cohort. Given the education level of the group, and the fact that they were drawn from a university chemistry department, we assumed that mouse familiarity was high.

|  | 1 | 2 | 3 | 4 | 5 |
|---|---|---|---|---|---|
| Prior VR Experience | 20 (62.5%) | 8 (25.0%) | 3 (9.38%) | 1 (3.13%) | 0 |
| Prior Tablet Experience | 1 (3.13%) | 5 (15.6%) | 9 (28.1%) | 10 (31.2%) | 7 (21.9%) |

Table S1: Self-reported familiarity with the VR and tablet platforms on a Likert scale for the first user study (n = 32), where one is no experience and five is very experienced. Mouse familiarity was assumed to be high.

*3.2 Second User Study*

In this study, 12 people were recruited and interviewed afterwards. Here we utilized a smaller sample size, because our emphasis during these studies was on gaining qualitative user feedback on attitudes to the three platforms, achieved via interview. Task accomplishment rates for this study are presented in Figure S2. In addition to carrying out questionnaires utilized during the first user study, thematic analysis (*5*) was performed on the recorded interview transcripts, a summary of which is presented below.

Participants were recruited in group sizes varying between one and three. In order to mitigate any learning or fatigue effect, the platform on which participants started was randomized. Specifically, 4 people started with the mouse platform, 4 people started with touchscreen platform and 4 started with the VR platform. Participants were rotated using a Latin Square in the order of VR, mouse, and touchscreen. Again, participants were first given the buckminsterfullerene task so that they could familiarize themselves with the interactive 'feel'. Before starting on a platform, participants were first shown a short, instructional video of the specific trefoil knot which they were being asked to tie. Once they grasped what they were being asked to do, participants were moved onto the knot tying task. A more detailed description of what both tasks entailed can be found below. Once the task was completed (or once time had elapsed), the process was repeated on the next platform until the participant had tried the task on all three platforms.

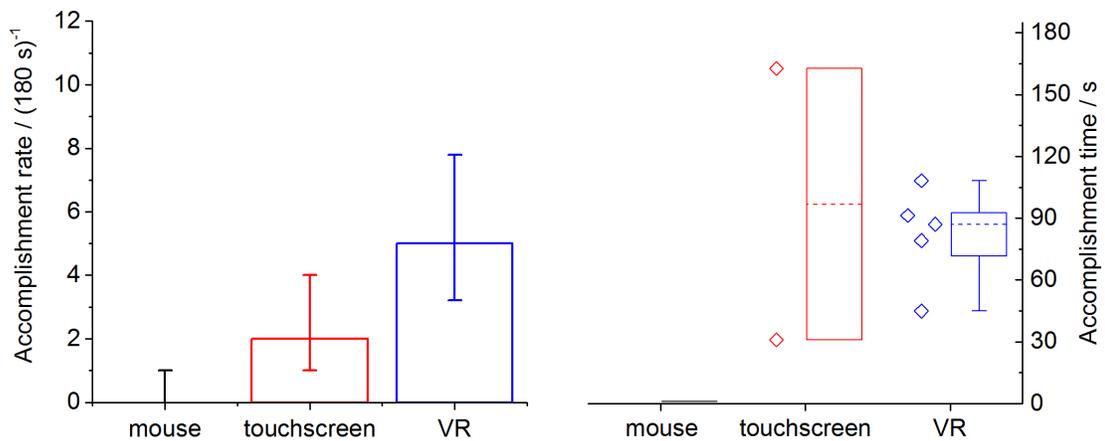

Figure S2: User study results. Left-hand panel shows user accomplishment rates for the knot tying task in the second user study (n = 12), with Poisson errors bars. Right hand panel shows the corresponding distribution of task accomplishment times, along with box-and-whisker plots, where whiskers span the data range, box limits indicate the standard error of the distribution, solid lines show the mean, and dotted lines show the median.

After each group had attempted the knot-tying task on each of the three platforms, they were sat down and interviewed about their experience using each of the four tasks. During this interview stage, the following points were covered:

(1) How had the participants found the task in general?
(2) Was there a preferred platform (or platforms) for completing the task?
(3) Why did participants prefer a given platform over others?
(4) Was there a least preferred platform (or platforms) for completing the task?
(5) Are there any suggestions for how the platforms can be improved?
(6) Any further points?

Again, participants for this study were recruited by emails to staff and students at the University of Bristol, and offered a £10 Amazon gift voucher for their time. Six (50%) of the participants were ages 18 to 24, five (41.7%) were ages 25 to 34, and one (8.3%) was age 35 to 44. Seven (58.3%) of participants were male. Participants reported a range of education levels. Four (33.3%) reported themselves as being undergraduate students, three (25%) reported themselves as being postgraduate students, four (33.3%) reported themselves as being post-doctoral researchers, and one (8.3%) reported themselves as being a research technician.

Participants were given a Likert scales question to complete in order to clarify their familiarity with using VR and tablets, where 1 represents having no experience and 5 represents being very experienced. A breakdown of responses can be found in Table S2. Altogether, self-reported VR experience was found to be low, where tablet use was more prevalent amongst the cohort. Given the education level of the group, mouse familiarity was again assumed to be high.

|  | 1 | 2 | 3 | 4 | 5 |
|---|---|---|---|---|---|
| Virtual Reality Experience | 7 (58.3%) | 4 (33.3%) | 0 | 1 (8.3%) | 0 |
| Tablet Experience | 1 (8.3%) | 0 | 6 (50%) | 2 (16.7%) | 3 (25%) |

Table S2: Self-reported familiarity with the VR and tablet platforms on a Likert scale for the second user study (n = 12), where one is no experience and five is very experienced. Mouse familiarity was assumed to be high.

*3.3 Third User Study*

We decided to repeat the methodology from the second user study with a larger sample size of 32 participants, identical to the sample sizes selected for the nanotube and helicene talks in the first user study. The primary aim of this leg was simply to obtain better statistics on knot task completion; therefore, no questionnaire or interview was given afterwards.

Participants were recruited in group sizes varying between one and three. In order to mitigate any learning or fatigue effect, the platform on which participants started was randomly selected. Specifically, 10 participants started with the mouse platform, 11 participants started with the touchscreen platform, and 11 participants started with the VR platform. Participants were rotated using a Latin Square in the order of VR, mouse, touchscreen.

Participants were recruited during the 5[th] annual UK CCPBioSim conference (13 – 14 Sept 2017), held at the University of Southampton. The chance to participate in the user study was advertised by email, flyers, and word of mouth. Six (18.75%) participants were ages 18 to 24, 20 (62.5%) reported themselves as being ages 25 to 34 years old, four (12.5%) reported themselves as being ages 35 to 44, one (3.125%) reported themselves as being ages 45 to 54, and one (3.125%) reported themselves as being over the age of 65. 24 (75%) of the participants were male. One (3.125%) participant chose not to state their gender. Participants reported a range of education levels. 19 (59.375%) participants were postgraduate students, 11 (34.375%) participants were postdoctoral researchers, and 2 (6.25%) of participants reported themselves as researchers.

Participants were given a Likert scales question to complete in order to ascertain their familiarity with using VR and tablets, where 1 represents having no experience and 5 represents being very experienced. A breakdown of responses can be found in Table S3. Self-reported VR experience was found to be low; tablet use was far more prevalent amongst the cohort. Given the education level of the group, mouse familiarity was again assumed to be high.

|  | 1 | 2 | 3 | 4 | 5 |
|---|---|---|---|---|---|
| Virtual Reality Experience | 20 (62.5%) | 7 (21.875%) | 3 (9.375%) | 0 | 2 (6.25%) |
| Tablet Experience | 1 (3.125%) | 3 (9.375%) | 10 (31.25%) | 12 (37.5%) | 6 (18.75%) |

Table S3: Self-reported familiarity with the VR and tablet platforms on a Likert scale for the third user studies (n = 32), where one is no experience and five is very experienced. Mouse familiarity was assumed to be high.

*3.4 Outline of User Tasks*

*Introductory buckminsterfullerene task*: Two buckminsterfullerene molecules were loaded into the Nano Simbox environment. Users were then instructed to grab the molecule and experiment with moving it around within the virtual space, and also to familiarize themselves with resizing and rotating the virtual space, thus giving them an understanding of the platform controls and the 'feel' of the molecular interaction. There was no time limit to this task, nor did it have any specific end goal. Section 1 of this SI includes the simulation details.

*Nanotube/methane task*: One $C_{60}$ nanotube and one methane molecule was loaded into the Nano Simbox environment. Users were instructed to grab the methane molecule and lead it through the centre of the nanotube, from one end to the other. The task had a time limit of 180 seconds and completion was marked as the point that the user had successfully pulled the methane molecule through the entire nanotube, i.e., leaving the opposite side from which it entered. Section 1 of this SI includes the simulation details.

*Helicene Task*: One 12-helicene molecule was loaded into the Nano Simbox environment. Users were instructed to manipulate the helicene molecule so that the screw-sense of the helix was reversed. The task had a time limit of 210 seconds and completion was marked as the point at which the screw-sense of the helicene molecule had gone from its clockwise initial conditions to being completely anti-clockwise. Section 1 of this SI includes the simulation details.

*Protein knot tying task*: One 17-alanine molecule was loaded into the Nano Simbox environment. Before starting the task on each of the platforms, users were shown a brief instructional video on the task. Users were then instructed to tie a $3_1$ or trefoil knot in 17-alanine. The task had a time limit of 180 seconds and completion was marked as the point at which a stable trefoil knot was formed in 17-alanine. The simulation details are specified above. Section 1 of this SI includes the simulation details.

*3.5 Platform controls*

Supplementary videos 1 and 2 give a good indication of how interaction in VR works, along with the controls which participants can access in order to accomplish the tasks outlined in the main text. We briefly outline those controls in what follows.

*Virtual Reality:* The simplest way for a user to change the perspective or angle from which they view a real-time molecular simulation is to walk around the molecule, exactly as they might do with a physical model. Alternatively, he/she has the option to rotate the 3d camera view by holding down the two 'grip' buttons on each side of the HTC vive controller (illustrated as purple rectangles in Fig S3a), and then carrying out the 3d rotation which he/she wishes to achieve. Scaling the size of the simulation render is achieved by holding down the two grip buttons, and moving the controllers either toward each other (zooming out from the simulated molecules) or away from each other (zooming in on the simulated molecules). To accomplish tasks, users pulled the triggers of their wireless controllers (shown in Fig S3a) to 'grab' specific atoms within the simulation, and then manipulate them as they like. The small purple sphere (situated at the end of the long axis of the controllers in Fig S3a and S3b) indicates the center of the Gaussian defined in Eq (4). Embedding the sphere into a particular atom and pulling the trigger results in a force being exerted on that particular atom, during which time the atom changes color. A menu button, shown in Fig S3b, is accessible by pressing the orange controller button. ↻ resets a particular simulation to its initial conditions, ❚❚ pauses the simulation, ▶❙ advances the simulation one frame at a time, and ▶ resumes the simulation at the standard propagation rate of 30 Hz. Additional menu options allow users to change the molecular renderer; however, these options were disabled during the user studies, in order to ensure that the renderers utilized on each platform were in fact identical.

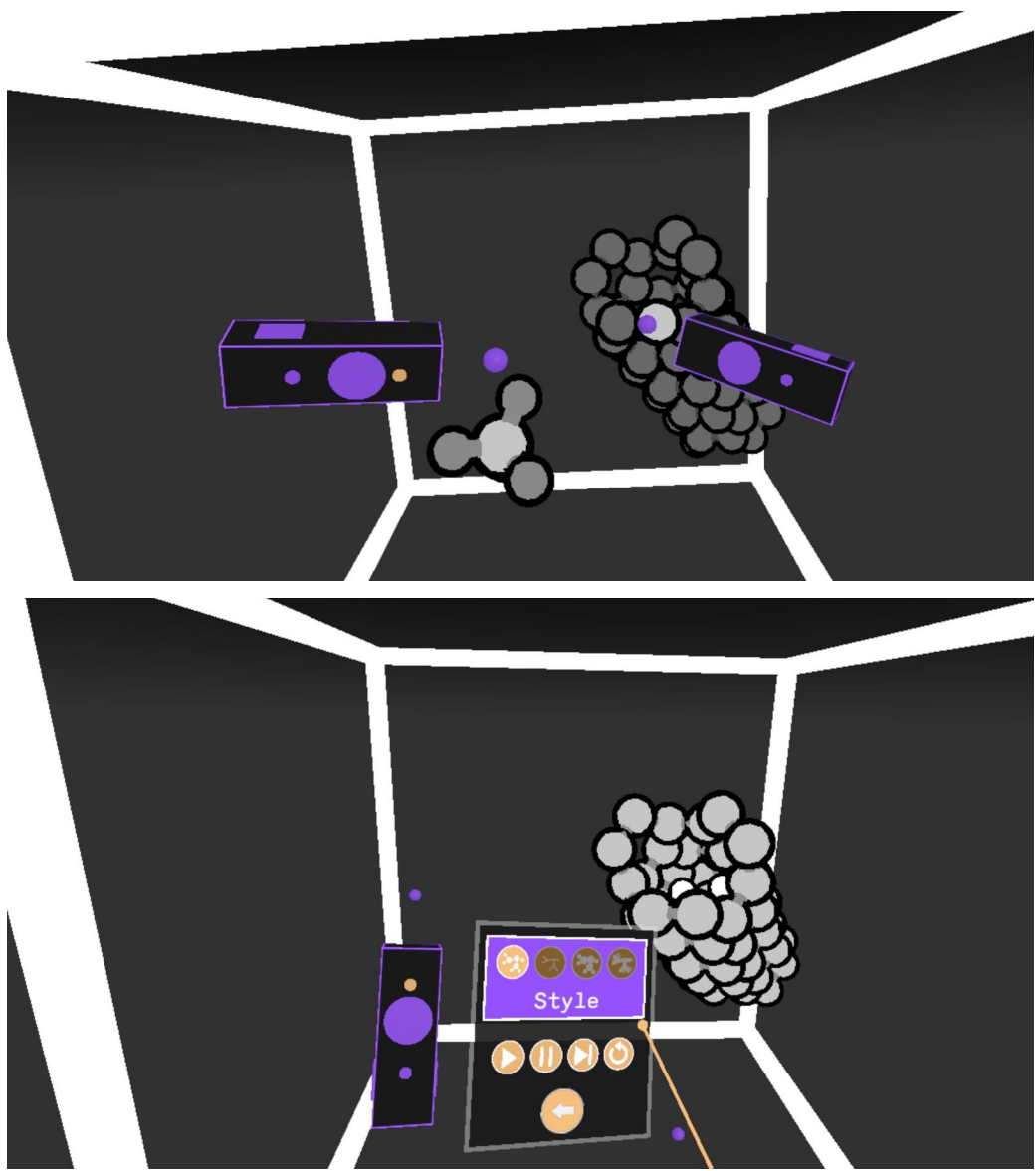

Figure S3: Screen shots of a user's view from within VR carrying out the nanotube task. The top panel (a) shows the user's controllers, along with the menu button, grip buttons, and positioning sphere. The bottom panel (b) shows the menu available when a user presses the menu button.

*Screen & Mouse:* When using a mouse to accomplish simulation tasks, the user is obviously unable to navigate the molecular simulation by walking around. Changing the camera view is therefore the only reorientation strategy. To change the rotation angles which define the camera's perspective with respect to the simulation box (see Fig S4a), we used the same strategy common to a number of molecular simulation/visualization tools, where the rotation angles of the camera in the vertical and horizontal directions can be changed by simply clicking the mouse, moving it across the screen, and then releasing once the desired rotational perspective has been achieved. To achieve such rotations, the user must ensure that the initial click is not located on an atom; in cases where the initial click occurs on an atom, then that particular atom is selected and 'grabbed', an interaction which persists until the user releases the mouse button. As an indicator that a user is 'grabbing' an atom, all atoms change color except for the atom being manipulated; once the atom is released, its color changes back to the default (this choice of visual feedback ensured consistency with the touchscreen interface, for which the design rationale is discussed below). As a result of the 2d limitations of the hardware, the user 'grabs' that atom measured to be nearest to the mouse icon in the 2d plane of the screen. Because atoms which are closer to the user are bigger, they are therefore easier to 'grab'. Once an atom is 'grabbed' it can only be moved in the 2d perspective of the camera; 3d motions must be built up from successive 2d motions followed by repositioning of the camera. To scale the size of the simulation render, we again used a strategy which is common to other molecular simulation/visualization programs, in which the user controls the zoom level via the mouse scroll wheel, with up/down scrolling motions allowing participants to zoom in/out. Fig S4b shows a user's view having undertaken a rotation, and a then subsequently zoomed in. The left hand panel of Figs S4a and S4b show the location of the ↺, ‖, ▶|, and ▶ icons, each of which has identical functionality to the VR version of the app.

*Touchscreen:* The touchscreen interface looks identical to the screen/mouse platform interface in Fig S4a and S4b. The only differences are in how interaction and navigation are carried out: (1) the user changes the camera orientation by moving a single finger across the touchscreen surface, ensuring that their initial contact position does not overlap with an atom; (2) zooming in an out is accomplished by the now ubiquitous 'pinch-to-zoom' gesture which is commonplace on mobile devices and touchscreen apps; (3) the user 'grabs' an atom by simply placing his/her finger on top of the atom that he/she wishes to manipulate. To indicate that a user is 'grabbing' an atom – i.e., exerting a force on a particular atom, he/she sees all atoms change color except for the atom which is being manipulated; once the atom is released, all colors change back to the default. This form of visual feedback we found to be particularly effective as a solution to the fact that the atom a user is trying to manipulate is often hidden under his/her finger; therefore, we required a form of visual feedback which could be easily perceived despite a finger obstructing the target.

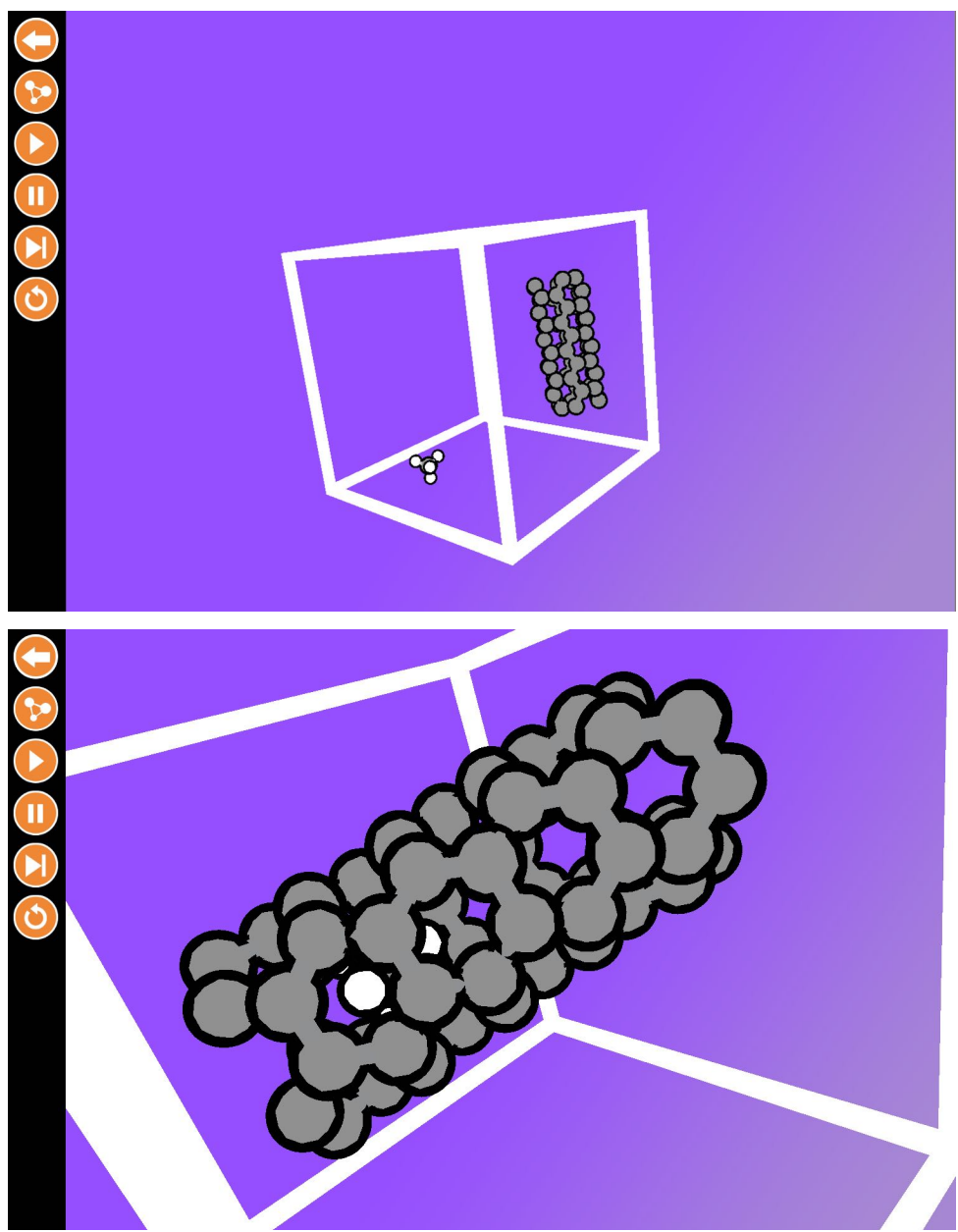

Figure S4: User's view of the molecular manipulation app when using either a mouse or touchscreen. Top panel (a) and bottom panel (b) shows the view from two different camera orientations and zoom levels.

*3.5 Thematic Analysis*

Interactively chaperoning molecular processes requires some degree of spatial reasoning: guiding chemical processes requires users to move, rotate and align molecules in 3D space. In order to gain a deeper understanding of the relationship between platform and spatial reasoning, participants were either given a questionnaire or interview after attempting tasks. Specifically, the 32 participants who completed the nanotube/methane and helicene tasks were given a questionnaire to complete. The 12 participants who completed the knot task were interviewed in groups of two or three in order to facilitate discussion and elicit more feedback. The task completion rates for first and third user tests can be found in the main text. Completion rates for the second user test can be found in Fig S2 within this SI.

Regardless of session, all participants were asked to state their favorite platform to use. Overwhelmingly, VR was selected as the preferred platform (41 out of 44 total participants). Commenting on the positive or negative features of VR, participants felt that it gave a distinct advantage to depth perception, making the 3D shape of molecules easily comprehendible. Inversely, as the other two platforms render molecules onto a flat screen, participants found it harder to see the overall shape of a molecule. Additionally, this attitude is corroborated by the questionnaire responses undertaken by the participants who completed the nanotube and helicene tasks. Participants were asked to rate the importance to task completion of depth perception, two-handed control, and orientating around the virtual space. Figure S5 provides a summary of these results. When asked to rate the importance of depth perception to task completion, all responses indicated it was important to some degree, by far the most positive response.

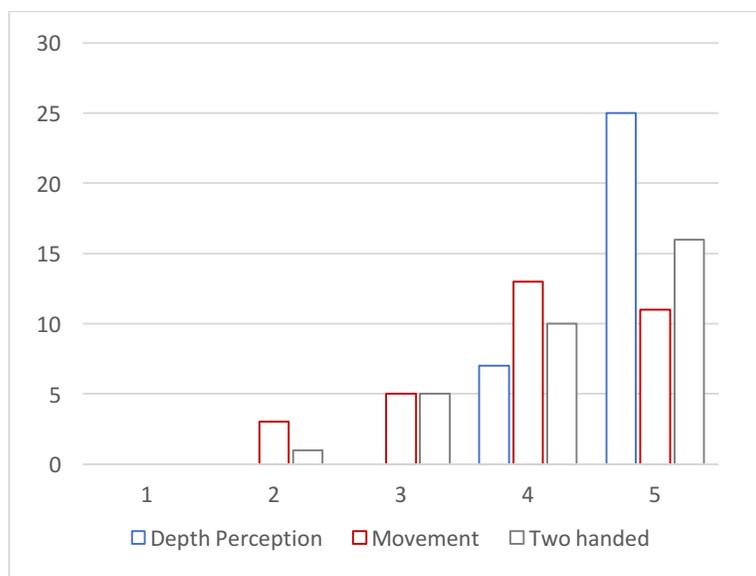

Figure S5: Participants self-reported attitudes to the importance of depth perception, navigating the virtual space, and controlling molecules with two hands. Responses were given on a Likert scale between 1 and 5, where 1 represents no importance and 5 represents high importance.

VR also had a second advantage with regards to depth perception. As the user is immersed in a fully-3D space, they can easily make movements into and out of the plane of their screen. When limited to the flat screen however – as is the case with the mouse and touchscreen platforms – participants found pinpointing 3D space more difficult. In particular, the two dimensional view caused difficulty for the knot task; participants needed to successfully circle one end of the protein string around the other and pull it through the resulting loop. With the three-dimensional 17-ALA molecule projected onto a two-dimensional screen, users found themselves changing 2d camera orientation in order to inspect whether the initial loop shape was appropriate for tying the knot. At this point, participants forfeited some degree of control over the molecule in order to adjust the camera; in several cases, the real-time molecular dynamics simulation led to a loss of molecular shape during the time that the user was adjusting the 2d camera view. In VR, not only can the camera and molecules be easily controlled simultaneously, users can also freely make molecular manipulations in any dimension, allowing a significantly higher degree of control.

As an additional consideration, the touchscreen camera and molecule were both controlled with the same action of dragging their finger against the screen, meaning that some participants would unintentionally rotate the camera at points where they may have intended to grab a molecule (and vice versa). In VR, camera view is controlled by head position, and the molecules are separately controlled by hand gestures, which removes input ambiguity. VR, as a co-located from of 3d interaction, directly maps user movements to in-world gestures, meaning that actions (e.g., tying a knot, or threading a methane molecule through a nanotube) were more intuitive. Participants expressed a sense of agency over the VR simulations owing to the fact that 'real-world' physical gestures can be directly utilized for VR molecular interactions. For example, the physical gesture required to tie a virtual molecular knot is essentially identical to that of tying a physical knot. For the two other platforms, this is not the case: participants are forced to translate what are familiar physical gestures (e.g., tying a knot) into a secondary set of gestures adapted to the limitations of the platform. In many cases, these secondary gestures are far less intuitive, especially when attempting to accomplish complicated 3d tasks.

## 4. Web URLs for Supplementary Videos

**Supplementary Video 1** is available at https://vimeo.com/244670465

**Supplementary Video 2** is available at https://vimeo.com/235894288